\documentclass[12pt,twoside]{article}
\usepackage{amsmath,amsfonts,amssymb,amsthm}
\usepackage{times}
\usepackage{hyperref}
\usepackage{graphicx}
\usepackage{stmaryrd}
\usepackage{color}

\newcommand{\be}{\begin{equation}}
\newcommand{\ee}{\end{equation}}
\newcommand{\bea}{\begin{eqnarray}}
\newcommand{\eea}{\end{eqnarray}}

\def\d_Vphi{\text{d}_V\hspace{-0.06em}\phi}
\def\d_Vphibar{\text{d}_V\hspace{-0.06em}\bar\phi}
\def\d_Vxi{\text{d}_V\hspace{-0.06em}\xi}

\def\be{\begin{eqnarray}}
\def\ee{\end{eqnarray}}
\def\beann{\begin{eqnarray*}}
\def\eeann{\end{eqnarray*}}
\def\beq{\begin{equation}}
\def\eeq{\end{equation}}
\def\ba{\begin{array}}
\def\ea{\end{array}}
\def\ben{\begin{enumerate}}
\def\een{\end{enumerate}}
\def\bea{\begin{eqnarray}}
\def\eea{\end{eqnarray}}

\def\5{\bar }
\def\6{\partial }
\def\7{\hat }
\def\4{\tilde }
\advance\textheight\topskip
\renewcommand{\tilde}{\widetilde}
\renewcommand{\hat}{\widehat}

\renewcommand{\d}{\partial}

\renewcommand{\geq}{\,{\geqslant}\,}

\newcommand{\binner}[2]{%
  {\langle}\kern-4.15pt{\langle}#1{,}\,#2{\rangle}\kern-4.15pt{\rangle}}

\newcommand{\ffrac}[2]{\raisebox{.5pt}%
  {\footnotesize$\displaystyle\frac{#1}{#2}$}\kern1pt}

\def\cL{\mathcal{L}}

\def\cY{\mathcal{Y}}

\numberwithin{equation}{section} \makeatletter
\@addtoreset{equation}{section}

\DeclareFontFamily{OT1}{rsfs}{} \DeclareFontShape{OT1}{rsfs}{m}{n}{
<-7> rsfs5 <7-10> rsfs7 <10-> rsfs10}{}
\DeclareMathAlphabet{\mycal}{OT1}{rsfs}{m}{n}

\newcommand*\xbar[1]{%
  \hbox{%
    \vbox{%
      \hrule height 0.5pt 
      \kern0.3ex
      \hbox{%
        \kern-0.0em
        \ensuremath{#1}%
        \kern-0.0em
      }%
    }%
  }%
} 

\begin{document}

\author{Glenn Barnich and Pierre-Henry Lambert}

\title{Einstein-Yang-Mills theory : \\ I. Asymptotic symmetries}

\date{}

\def\mytitle{Einstein-Yang-Mills theory : \\ I. Asymptotic symmetries}

\pagestyle{myheadings} \markboth{\textsc{\small G.~Barnich, P.-H.~Lambert}}{%
  \textsc{\small Einstein-Yang-Mills asymptotic symmetries}}
\addtolength{\headsep}{4pt}


\begin{centering}

  \vspace{1cm}

  \textbf{\Large{\mytitle}}


  \vspace{1.5cm}

  {\large Glenn Barnich and Pierre-Henry Lambert}

\vspace{.5cm}

\begin{minipage}{.9\textwidth}\small \it  \begin{center}
   Physique Th\'eorique et Math\'ematique \\ Universit\'e Libre de
   Bruxelles and International Solvay Institutes \\ Campus
   Plaine C.P. 231, B-1050 Bruxelles, Belgium
 \end{center}
\end{minipage}

\end{centering}

\vspace{1cm}

\begin{center}
  \begin{minipage}{.9\textwidth}
    \textsc{Abstract} Asymptotic symmetries of the Einstein-Yang-Mills
    system with or without cosmological constant are explicitly worked
    out in a unified manner. In agreement with a recent conjecture,
    one finds a Virasoro-Kac-Moody type algebra not only in three
    dimensions but also in the four dimensional asymptotically flat
    case.
  \end{minipage}
\end{center}

\thispagestyle{empty}

\vfill

\newpage

 \tableofcontents

\section{Introduction}
\label{sec:introduction}

Even though the first discussions of asymptotic symmetries dealt with
four dimensional general relativity, both at null
\cite{Bondi:1962px,Sachs:1962wk,Sachs2} and at spatial infinity
\cite{Regge:1974zd}, most of the recent work was devoted to three
dimensions because of the occurence of a classical central charge
\cite{Brown:1986nw} that plays a key role in symmetry based
explanations \cite{Strominger:1998eq} of the entropy of the BTZ black
hole \cite{Banados:1992wn,Banados:1993gq} and in other aspects of the
AdS/CFT correspondence (see e.g.~\cite{Aharony:1999ti}, chapter 5).

In recent work \cite{Strominger:2013lka}, Strominger suggested to
extend the analysis for gravity in four dimensions at null infinity to
include Yang-Mills fields and established a relation to field
theoretic soft photon and graviton theorems
\cite{PhysRev.140.B516}. During these considerations, the symmetry
algebra was argued to be of Virasoro-Kac-Moody type.

In this note, we confirm this conjecture. We start by showing that the
residual symmetry algebra of a standard gauge choice adapted to the
asymptotic analysis of the Einstein-Yang-Mills system is simply the
gauge algebra in one dimension lower. The asymptotic symmetry algebra
is then obtained by a further reduction that comes from suitable
fall-off conditions on the remaining fields. Details for various
standard cases, including the flat case with asymptotics at null
infinity, are provided.

At this stage, one might wonder why the enhancement of the $U(1)$
electromagnetic gauge symmetry has not been discussed in previous
detailed investigations of the asymptotic properties of the
Einstein-Maxwell system
\cite{janis:902,vandeBurg06051969,Exton:1969im}. With hindsight, the
reason is that the focus was on the modifications of the equations of
motions and their solutions due to the presence of the
electro-magnetic field which had been included through its field
strength. It turns out however that a formulation in terms of gauge
potentials is required if one wants to discuss action principles and
asymptotic symmetries for both gravitational and Yang-Mills type gauge
fields in a unified manner.

With the symmetries under control, the next stage is to work out
asymptotic solutions. This should be done, for simplicity first in
three, and then in four dimensions, along the lines of the detailed
analysis of the Einstein-Maxwell system. Once this is done, the
symmetry transformations of the fields characterizing asymptotic
solutions can be computed. Then one is ready to work out the
holographic current algebra, including potential central
extensions. These questions will be adressed elsewhere.

\section{Gauge structure of the Einstein-Yang-Mills system}

The Einstein-Yang-Mills system in $d$ dimensions is described by the
action 
\begin{equation}
  \label{eq:1}
  S=\frac{1}{16\pi G}\int d^dx \sqrt{|g|}~~
  [R-2\Lambda-g_{ij}F^i_{\mu\nu}F^{j\mu\nu}], \quad \Lambda=-\frac{(d-1)(d-2)}{2l^2},
\end{equation}
where $g_{ij}$ is an invariant non-degenerate metric in a basis $T_i$
of the internal gauge algebra $\mathfrak{g}$, $F_{\mu\nu}=\partial_\mu
A_\nu-\partial_\nu A_\mu+[A_\mu,A_\nu]$ is the field strength,
$A_\mu=A^i_\mu T_i$ and the bracket denotes the Lie bracket in
$\mathfrak{g}$, $[T_i,T_j]=f^k_{ij}T_k$. 

The complete gauge algebra consists of pairs $(\xi,\epsilon)$ of a
vector field $\xi^\mu\d_\mu$ and an internal gauge parameter
$\epsilon^i T_i$. A generating set of gauge symmetries is given by
\begin{equation}
  \label{eq:5}
  \delta_{(\xi,\epsilon)} g_{\mu\nu}=-\cL_\xi
g_{\mu\nu},\quad \delta_{(\xi,\epsilon)}A_\mu=-\cL_\xi A_\mu
+D^A_\mu\epsilon,
\end{equation}
with $D_\mu^A\epsilon=\d_\mu\epsilon+[A_\mu,\epsilon]$. 

Let the fields be collectively denoted by
$\phi^\alpha=(g_{\mu\nu},A_\mu)$.  When the gauge parameters
$(\xi,\epsilon)$ depend only on the spacetime coordinates but not on
the fields $\phi^\alpha$, one has 
\begin{equation}
[\delta_{(\xi_1,\epsilon_1)},\delta_{(\xi_2,\epsilon_2)}]\phi^\alpha
=\delta_{(\hat\xi,\hat\epsilon)}\phi^\alpha,
\end{equation}
with $\hat\xi=[\xi_1,\xi_2]$ the Lie bracket for vector fields and
$\hat\epsilon=\xi^\mu_1\d_\mu\epsilon_2-\xi^\mu_2\d_\mu\epsilon_1
+[\epsilon_1,\epsilon_2]$. The Lie bracket for field independent gauge
parameters is given by
\begin{equation}
[(\xi_1,\epsilon_1),(\xi_2,\epsilon_2)]=
  (\hat\xi,\hat\epsilon)\label{eq:6a}.
\end{equation}

In the case of gauge parameters $(\xi,\epsilon$) that are field
dependent, one finds instead 
\begin{equation}
[\delta_{(\xi_1,\epsilon_1)},\delta_{(\xi_2,\epsilon_2)}]\phi^\alpha
=\delta_{(\hat\xi_M,\hat\epsilon_M)}\phi^\alpha,
\end{equation}
with 
\begin{align}
  \hat\xi_M&=\hat\xi+\delta_{(\xi_1,\epsilon_1)}\xi_2-
  \delta_{(\xi_2,\epsilon_2)}\xi_1,\label{eq:103}\\
  \hat\epsilon_M&=\hat\epsilon+\delta_{(\xi_1,\epsilon_1)}
  \epsilon_2-\delta_{(\xi_2,\epsilon_2)}\epsilon_1,\label{eq:104}
\end{align}
and the Lie (algebroid) bracket for field dependent gauge parameters is thus
defined through
\begin{eqnarray}\label{eq:105}
[(\xi_1,\epsilon_1),(\xi_2,\epsilon_2)]_M=(\hat\xi_M,\hat\epsilon_M).
\end{eqnarray}

\section{Dimensional reduction through gauge fixation}
\label{sec:dimens-reduct-thro}

In terms of coordinates $x^\mu=(u,r,x^A)$, where $x^A$ are angular
variables in $d-2$ dimensions, we make the following gauge fixing
ansatz for the metric and Yang-Mills potentials:
\begin{equation}
\begin{split}
g_{\mu\nu}&=
\begin{pmatrix}
e^{2\beta}\dfrac{V}{r}+g_{CD} U^CU^D&	-e^{2\beta}	&	-g_{BC}U^C\\
-e^{2\beta}	&	0	&	0\\
-g_{AC}U^C&	0	&g_{AB}
\end{pmatrix},\label{eq:2.7}\\
A_\mu&=
\begin{pmatrix}
A_u,0,A_A
\end{pmatrix}.
\end{split}
\end{equation}
In addition, one imposes the determinant condition ${\rm det}\
g_{AB}=r^{2(d-2)}{\rm det}\bar\gamma_{AB}$, with $\bar\gamma_{AB}$ the
metric on the unit $d-2$-sphere. 

As in the purely gravitational four dimensional case
\cite{Sachs:1962wk}, these conditions fix the gauge freedom up to some
$r$ independent functions. Indeed, the gauge transformations
\eqref{eq:5} that preserve this gauge choice, i.e., the residual gauge
symmetries, are determined by gauge parameters that have to satisfy
\begin{equation}
\cL_\xi g_{rr}=0,\hspace{0.5cm}\cL_\xi g_{rA}=0,
\hspace{0.5cm}g^{AB}\cL_\xi g_{AB}=0,\hspace{0.5cm}
-\cL_\xi A_{r}+D^A_r \epsilon=0.\label{eq:2.9}
\end{equation}
This gives the differential conditions
\begin{equation}
  \label{eq:7}
\begin{split}
 & \d_r\xi^u=0,\quad \d_r\xi^A=\d_B\xi^u g^{AB}e^{2\beta},\\ & \d_r
  (\frac{\xi^r}{r})=-\frac{1}{d-2} (\bar D_B \d_r \xi^B-\d_B\xi^u\d_r
  U^B),\quad \d_r \epsilon=\d_B\xi^u g^{AB}e^{2\beta} A_A, 
\end{split}
\end{equation}
the general solution of which is
\begin{equation}
  \begin{split}
& \xi^u =F(u,x^A),\quad \xi^A=Y^A(u,x^B)-\d_B F\int^\infty_r dr^\prime (e^{2\beta}
  g^{AB}),  \\ & \xi^r=-\frac{r}{d-2} (\bar D_B \xi^B-\d_B F
  U^B),\quad
\epsilon = E(u,x^A) - \d_B F\int^\infty_r dr^\prime
  (g^{BA}e^{2\beta} A_A),\label{eq:2.10}
  \end{split}
\end{equation}
and involves $d-1+n$ arbitrary $r$-independent functions
$F(u,x^A),Y^A(u,x^B),E^i(u,x^A)$. 

At this stage, it is sufficient to impose the following fall-off
conditions on the components of the metric and the gauge potentials,
\begin{equation}
  \label{eq:6}
 e^{2\beta}g^{AB}=O(r^{-1-\epsilon})=e^{2\beta}g^{AB}A_A,\quad \ U^C
 e^{2\beta}g^{AB} =o(r^{-1})\ {\rm
   for}\ d>3.
\end{equation}
In particular, the first of these conditions guarantee that the
integrals for $\xi^A$ and $\epsilon$ in \eqref{eq:2.10} are
well-defined and that $\lim_{r\to\infty}\xi^A=Y^A$,
$\lim_{r\to\infty}\epsilon=E$. 

Consider then the vector fields
$\xi^R=F\partial_u+Y^A\partial_A$ and the internal gauge parameter
$\epsilon^R=E^iT_i$, equipped with the Lie bracket
\begin{equation}
[(\xi^R_1,\epsilon^R_1),(\xi^R_2,\epsilon^R_2)]
=(\hat\xi^R,\hat\epsilon^R),\label{eq:7a}
\end{equation}
for field independent gauge parameters \eqref{eq:6a} of the
Einstein-Yang-Mills system in $d-1$ dimensions. 
We are now ready to state the main result of this section: 

{\em The Lie algebra of residual gauge parameters \eqref{eq:2.10}
  equipped with the Lie bracket $[\cdot,\cdot]_M$ of the $d$
  dimensional Einstein-Yang-Mills system is a faithful
  representation of the Lie algebra of field independent gauge
  parameters $(\xi^R,\epsilon^R)$ of the $d-1$ dimensional
  Einstein-Yang-Mills system.}

The proof for the diffeomorphism part is almost exactly the same as in
\cite{Barnich:2010eb}, except for the additional $u$ dependence in
$Y^A$, which is easily taken into account. We will thus not repeat all
details here. First, one needs to check that the result holds for
$\hat\xi^u_M$, $\hat\xi^A_M$, $r^{-1}\hat\xi^r_M$, $\hat\epsilon_M$
at $r\to\infty$. This is where the fall-off conditions \eqref{eq:6}
are needed. Note however that the fall-off condition on $U^A$ have
been considerably relaxed and, in particular, there are no conditions
for $d=3$. The rest of the proof consists in verifying that $\d_r
\hat\xi^u_M$, $\d_r \hat\xi^A_M$, $\d_r(r^{-1}\hat\xi^r_M)$,
$\d_r\hat\epsilon_M$ satisfy equations \eqref{eq:7} with
$(\xi^R,\epsilon^R)$ replaced by $(\hat\xi^R,\hat\epsilon^R)$.

\section{Fall-off conditions and asymptotic symmetry structure}
\label{sec:fall-cond-asympt}

Suppose now that in spacetime dimensions $4$ or higher, precise
fall-off conditions for the metric coefficients and gauge potentials
are given by
\begin{equation}
\begin{split}
  & \beta=o(1),\quad U^A=o(1),\quad g_{AB}dx^A
  dx^B=r^2\bar\gamma_{AB}(x^C) dx^Adx^B+o(r^2), \\
  & \frac{V}{r}=-\frac{r^2}{l^2} +o(r^2),\quad A_u=o(1),\quad
  A_B=A^{0}_B(u,x^C)+o(1). \label{eq:2}
\end{split}
\end{equation}
In the asymptotically flat $U(1)$ case in four dimensions, these
fall-off conditions are consistent with those of
\cite{janis:902,vandeBurg06051969,Exton:1969im}. They imply in
particular the conditions required in \eqref{eq:6}.

The gauge transformations that preserve these fall-off conditions have
to satisfy, in addition to \eqref{eq:2.9}, the supplementary
conditions
\begin{equation}
\begin{split}
  &\cL_\xi g_{ur}=o(1),\hspace{0.5cm}\cL_\xi
  g_{uA}=o(r^2),\hspace{0.5cm}
  \cL_\xi g_{AB}=o(r^2),\hspace{0.5cm}\cL_\xi g_{uu}=o(r^2),\\
  &-\cL_\xi A_u+D^A_u \epsilon=o(1),\hspace{0.5cm} -\cL_\xi A_B+D^A_B
  \epsilon=O(1).\label{eq:2.21}
\end{split}
\end{equation}
They are equivalent to the following differential equations on the
$(\xi^R,\epsilon^R)$
\begin{equation}
\begin{split}
  \label{eq:2.22}
&  \d_u F=\frac{1}{d-2}\Psi, \quad \d_u
Y^A\bar\gamma_{AB}=\frac{1}{l^2} \d_B F, \\
& \cL_Y\bar\gamma_{AB} =
\frac{2}{d-2} \Psi \bar\gamma_{AB},\quad \d_u
E=\frac{1}{l^2}\d^B F A^{0}_B, 
\end{split}
\end{equation}
with $\Psi=\bar D_B Y^B$, the general solution of which is
\begin{equation}
\begin{split}
 & F =f(x^A)+\dfrac{1}{d-2}\int_0^u du^\prime~~\Psi,\quad
  Y^A=y^A(x^B)+\dfrac{1}{l^2}\int^u_0 du^\prime
  (\bar\gamma^{AB}\d_B F),  \\
 & E = e(x^A)+\dfrac{1}{l^2}\int^u_0 du^\prime(\bar\gamma^{AB}\d_A F
  A_B^0).\label{eq:4.5}
\end{split}
\end{equation}
Let us denote by $a_B$ the values of $A^{0}_B$ at $u=0$ and consider
time independent conformal Killing vectors of the $d-2$ sphere,
\begin{equation}
  \label{eq:4}
\cL_y \bar\gamma_{AB}=\frac{2}{d-2}\psi\bar\gamma_{AB},
\end{equation} 
with $\psi=\bar D_B y^B$. In addition, in the case of a non-vanishing
cosmological constant, the vectors $\d^Af=\bar\gamma^{AB}\d_Bf$ are
also required to be conformal Killing vectors of the $d-2$ sphere, as
follows by differentiating the third of \eqref{eq:2.22} with respect
to $u$ and setting $u=0$. A second derivative with respect to $u$ at
$u=0$ then implies that $\d^A\psi$ are also conformal Killing vectors
of the $d-2$ sphere. This can be continued for higher order
derivatives.

In terms of these quantities, the asymptotic symmetry structure is
described through the brackets
\begin{equation}
\begin{split}
  \label{eq:3}
  & \hat f=\frac{1}{d-2}f_1\psi_2+y^A_1\d_A f_2-(1\leftrightarrow
  2),\quad \hat y^A=\frac{1}{l^2} f_1 \d^A f_2+ y^B_1\d_B
  y^A_2-(1\leftrightarrow 2),\\
  & \hat e=\frac{1}{l^2} f_1 \d^A f_2 a_A +y^A_1\d_A
  e_2-(1\leftrightarrow 2)+[e_1,e_2].
\end{split}
\end{equation}
On account of the explicit field dependence in $\hat e$, one has to
use the Lie algebroid bracket in order to check the Jacobi identity
for $e$ with $\delta_{y,e} a_A=-\cL_y a_A+ D^a_A e$ in the case of a
non-vanishing cosmological constant. More generally, it is implicitly
understood that each time an element depends explicitly on the fields,
the Lie algebroid bracket has to be used.

By the same reasoning as before, one then shows that the asymptotic
symmetry structure is represented at infinity for all values of $u$
through the Lie algebroid bracket that involves the dependence on
$A^0_B$, and then also in the bulk spacetime through the result of the
previous section.

The gauge theory part of the asymptotic symmetry structure consists of
elements of the form$(0,0,e)$. It is is a non-abelian ideal that
contains an arbitrary $\mathfrak g$-valued function on the $d-2$
sphere. In that sense, it is a generalisation of a loop algebra where
the base space is a higher dimensional sphere instead of a circle.

The quotient of the total structure by this ideal is the spacetime
part. It can be described by elements of the form $(f,y^A,0)$ with
brackets determined by the first line of \eqref{eq:3}. In the case of
vanishing cosmological constant, elements of the form $(f,y^A,0)$ form
a subalgebra that acts on the gauge theory ideal.

\section{Explicit description of asymptotic symmetry structure in
  particular cases}
\label{sec:expl-descr-part}

\subsection{Dimensions 4 and higher, anti-de Sitter case}
\label{sec:dimensions-5-or}

For $d\geq 4$ and $l\neq 0$, the space-time part of the asymptotic
symmetry structure is isomorphic to $\mathfrak{so}(d-1,2)$, the algebra
of exact Killing vectors of $d$-dimensional anti-de Sitter space, in
agreement with the analysis in \cite{Henneaux:1985ey}. 

Indeed, in the coordinates we are using, the anti-de Sitter metric is
given by
\begin{equation}
g_{\mu\nu}=\begin{pmatrix}
-\frac{r^2}{l^2}-1&	-1	&	0\\
-1	&	0	&	0\\
0 &	0	&r^2\bar\gamma_{AB}
\end{pmatrix}.\label{eq:ads}
\end{equation} 
Besides the conditions 
\begin{gather}
  \label{eq:4a}
  \bar\xi^u=\bar F(u,x),\quad \bar\xi^A=\bar Y^A(u,x)-\frac{1}{r}\d^A
  \bar F,\quad
  \bar\xi^r=\frac{1}{d-2}(-r\bar \Psi+\bar \Delta \bar F),\\
  \d_u \bar F=\frac{1}{d-2}\bar \Psi,\quad \d_u \bar
  Y^A=\frac{1}{l^2}\d^A \bar F,
\end{gather}
where $\bar Y^A$ and $\d^A \bar F$ are conformal Killing vectors of
$\bar\gamma_{AB}$, which correspond to an asymptotic Killing vector
evaluated for the anti-de Sitter metric, an exact Killing vector
$\bar\xi=\bar\xi^u\d_u +\bar\xi^r\d_r +\bar \xi^A\d_A$ has also to
satisfy the additional conditions
\begin{equation}
  \label{eq:8}
  \d_B \bar F=-\frac{1}{d-2}\d_B\bar\Delta \bar F, \quad  
\bar \Psi=-\frac{1}{d-2}\bar\Delta \bar \Psi.
\end{equation}
The latter are automatically satisfied for conformal Killing vectors
$\bar Y^A,\d^A \bar F$ of the unit $d-2$ sphere.

Even though it is not needed for this proof, one can also check
directly that, if $y^A$ are conformal Killing vectors of the $d-2$
sphere, then the requirement that $\d^A\psi$ are also conformal
Killing vectors is automatically satisfied if $d\neq 4$, while for
$d=4$, this reduces the local conformal algebra in $2$ dimensions to
the globally well defined algebra $\mathfrak{so}(3,1)$ on the $2$
sphere.

\subsection{Dimensions 5 and higher, flat case}
\label{sec:dimensions-5-higher}

For $l\to\infty$, the asymptotic symmetry structure of field
independent parameters $(f,y^A,e)$ simplifies. The subalgebra
$(0,y^A,0)$ of conformal Killing vectors of the $d-2\geq 3$ sphere
represents the Lorentz algebra $\mathfrak{so}(d-1,1)$. It acts both on the
abelian ideal $(f,0,0)$ of arbitrary functions on the sphere, 
representing supertranslations, and on the gauge theory ideal. 

Stronger fall-off conditions motivated by the Einstein equations of
motions have been considered in \cite{Tanabe:2011es}. They require
$\d^A F$ to be conformal Killing of the $d-2$ sphere. In turn this
requires both $\d^A f$ and $\d^A\psi$ to be conformal Killing
vectors. Again, by comparing with the conditions satisfied by exact
Killing vectors of Minkowski space-time, the only additional
conditions are \eqref{eq:8}, which are automatically satisfied for
conformal Killing vectors $Y^A,\d^A F$. This shows that the additional
conditions reduce super to standard translations so that the spacetime
part of the asymptotic structure becomes the Poincar\'e algebra
$\mathfrak{iso}(d-1,1)$.

\subsection{4 dimensional flat case}
\label{sec:4-dimens-asympt}

In $4$ dimensions, it is useful to introduce stereographic coordinates
$\zeta=\cot{\frac{\theta}{2}}e^{i\phi}$ and its complex conjugate, so
that $\bar\gamma_{AB} dx^Adx^B=2P_S^{-2}d\zeta d\xbar\zeta$ with
$P_{S}=\frac{1}{\sqrt 2}(1+\zeta\xbar\zeta)$. The covariant derivative
on the $2$ surface is then encoded in the operator
\begin{equation}
\eth \eta^s= P^{1-s}_S\bar \d(P^s_S\eta^s),\qquad \xbar \eth
\eta^s=P^{1+s}\d(P^{-s}\eta^s)\,, 
\label{eq:34}
\end{equation}
where $\eth,\bar\eth$ raise respectively lower the spin weight $s$ by
one unit and satisfy
\begin{equation} 
[\bar \eth, \eth]\eta^s=\frac{s}{2}  R_S\,
  \eta^s\,,\label{eq:35}
\end{equation}
with $ R_S=4P^2_S\d \bar \d \ln P_S=2$. 

Let $\cY=P^{-1}_S y^{\bar\zeta}$ and $\xbar \cY=P^{-1}_S y^\zeta$ be of spin
weights $-1$ and $1$ respectively. The conformal Killing equations and
the conformal factor then become
\begin{equation}
  \label{eq:25}
  \eth \xbar \cY=0=\xbar\eth \cY,\qquad \psi=(\eth \cY+\xbar\eth
  \xbar \cY)\,.
\end{equation}
It follows for instance that $\xbar\eth \eth \cY=-\cY$, $\eth^2
\psi=\eth^3\cY$, $\xbar\eth\eth \psi=-\psi$. 

In order to describe the asymptotic symmetry structure there are then
two options. 

The first is to require well-defined functions on the $2$-sphere. This
amounts to restricting oneself to the conformal Killing vectors that
satisfy $\eth^3\cY=0=\xbar\eth^3\xbar\cY$ and require that the
functions $f,e^a$ that occur in \eqref{eq:3} (with $l\to\infty$) can
be expanded in spherical harmonics. The asymptotic symmetry algebra is
then the semi-direct sum of the globally well-defined
$\mathfrak{bms}^{\rm glob}_4$ algebra \cite{Bondi:1962px,Sachs2} with
a globally well-defined ``sphere'' algebra.

Alternatively \cite{Barnich:2009se,Barnich:2010eb}, one admits Laurent
series and expands $y^{\zeta}\d_\zeta$ in terms of
$l_m=-\zeta^{m+1}\d_\zeta$, $y^{\bar\zeta}\d_{\bar\zeta}$ in terms of
$\bar l_m$, $f$ in terms of $t_{m,n}=P_S^{-1}\zeta^m\bar\zeta^n$ and
$e$ in terms of $j_i^{m,n}=T_i\zeta^m\bar\zeta^n$. In these terms, the
non-vanishing brackets of the asymptotic symmetry algebra become
\begin{align}
[l_l,t_{m,n}]&=\left(\dfrac{l+1}{2}-m\right)t_{m+l,n},
\quad [l_m,l_n]=(m-n)~~l_{m+n},\\
[\bar l_l,t_{m,n}]&=\left(\dfrac{l+1}{2}-n\right)t_{m,n+l},
\quad [\bar l_m,\bar l_n]=(m-n)~~\bar l_{m+n},\\
[l_l,j^{m,n}_i,]&=-m~~j^{m+l,n}_i,\quad
[\bar l_l,j^{m,n}_i]=-n~~j^{m,n+l}_i,\\\label{eq:3.67}
[j^{l,p}_i,j^{m,n}_j]&= f_{ij}^k~~j_k^{l+m,p+n}. 
\end{align}

\subsection{3 dimensional anti-de Sitter case}
\label{sec:3-dimensional-anti}

On the metric components, we use the same fall-off conditions as in
\ref{sec:fall-cond-asympt}. Note that the determinant condition
requires $g_{\phi\phi}=r^2$ and that the fall-off conditions allow for
$\ln r$ terms both in $g_{uu}$ and $g_{u\phi}$. The spacetime part of
the asymptotic symmetry structure is then described by two copies of
the conformal algebra \cite{Brown:1986nw},
$F\d_u+Y\d_\phi=Y^+(x^+)\d_++Y^-(x^-)\d_-$, where
$x^\pm=\frac{u}{l}\pm \phi$.

In order to accommodate the charged and rotating black hole solution
\cite{Martinez:1999qi}, the fall off conditions on the gauge
potentials can be chosen as $A_+=O(\ln r)$, while one simultaneously
requires $A_-=o(1)$.  Alternatively, one could also exchange the
r\^ole of $+$ and $-$.  Requiring $-\cL_\xi A_+ +D^A_+ \epsilon=O(\ln
r)$ gives no conditions, while $-\cL_\xi A_- +D^A_- \epsilon=o(1)$
leads to $\d_-E=0$. In this case, there is no explicit field
dependence and the asymptotic symmetry structure simplifies as
compared to the higher dimensional case.

When expanding $Y^+\d_+,Y^-\d_-,E$ in terms of modes,
$l^{\pm}_m=e^{imx^\pm}\d_\pm,j^m_i=T_ie^{imx^+}$, the non-vanishing
brackets of the asymptotic symmetry algebra are explicitly given by
\begin{equation}
i[l^\pm_m,l^\pm_n]=(m-n) l^\pm_{m+n},\quad i[l^+_m,j^n_i]=-n
j^{m+n}_i,\quad i[j^m_i,j^n_i]=if_{ij}^k j^{m+n}_k. 
\end{equation}

\subsection{3 dimensional flat case}
\label{sec:3-dimensional-flat}

In this case, the spacetime part of the asymptotic symmetry structure
is the $\mathfrak{bms}_3$ algebra described by
$F\d_u+Y\d_\phi=[f(\phi)+uy(\phi)]\d_u+y\d_\phi$. For the gauge
potentials, one may then choose $A_u=o(1)$, $A_\phi=O(\ln r)$. 
Requiring $-\cL_\xi A_u+D_u^A \epsilon=o(1)$ leads to
$\d_u E=0$, while $-\cL_\xi A_\phi+D_\phi^A \epsilon=O(\ln r)$
gives no conditions. 

When expanding $F\d_u+Y\d_\phi,E$ in terms of modes,
$l_m=e^{im\phi}\d_\phi+uim\phi e^{im\phi}\d_u$, $t_m=e^{im\phi}\d_u$,
$j^m_i=T_i e^{im\phi}$, the non vanishing brackets of the asymptotic
symmetry algebra are given by
\begin{equation}
\begin{split}
  \label{eq:10}
 & i[l_m,l_n]=(m-n) l_{m+n},\quad i[l_m,t_n]=(m-n) t_{m+n},\\
 & i[l_m,j^n_i]=-n j^{m+n}_i,\quad i[j^m_i,j^n_j]=if_{ij}^k j^{m+n}_k.
\end{split}
\end{equation}

\section{Discussion}
\label{sec:discussion}

The gauge fixing and fall-off conditions that we have considered have
been mainly dictated by the desire to yield the usual asymptotic
symmetry structure, at least for the spacetime part, while otherwise
being as relaxed as possible. As partly already discussed in the text,
additional more restrictive conditions motivated by finiteness of
associated conserved currents or their integrability for example can
further reduce the asymptotic symmetry structure, in particular also
in the Yang-Mills part.

On the other hand, one may wonder how far these conditions can be
relaxed even further. From a holographic point of view, the role of
the gauge fixing conditions considered in section
\ref{sec:dimens-reduct-thro} is to fix the dependence in $r$ of the
gauge parameters. This can be achieved in various ways. In the
Newman-Unti gauge \cite{newman:891} for instance, one can require
$g_{ur}=-1$ instead of the determinant condition, leading to another
integration function in $\xi^r$, that may or may not be fixed through
additional conditions, see e.g.~\cite{Barnich:2011ty}. One may also relax
conditions \eqref{eq:6}. For instance at fixed but finite $r$ no such
conditions are needed. Even though one will then not get the symmetry
structure of the Einstein-Yang-Mills system in one dimension lower,
the resulting structure will still be well-defined.

Similarly, except for the fall-off conditions on $g_{AB}$, the role
of the other conditions in section \ref{sec:fall-cond-asympt} is to
fix the time dependence of the gauge parameters and thus of the
symmetry structure of the dual boundary theory. In the present set-up,
the fall-off conditions on $g_{AB}$ are the only ones that constrain
the dependence of the symmetry structure, or more precisely of
$Y^A,F$, on the spatial coordinates $x^A$. In other words, relaxing
this condition leads to ``superrotations'' that, like
supertranslations and the Yang-Mills gauge parameters, have an
arbitrary $x^A$ dependence.

\section*{Acknowledgements}
\label{sec:acknowledgements}

\addcontentsline{toc}{section}{Acknowledgments}

The authors thank L.~Donnay and C.~Troesssaert for useful discussions.
G.B.~is Research Director of the Fund for Scientific Research-FNRS
Belgium, P.-H.L.~benefits from a PhD fellowship of the ULB. This work
is supported in part by the Fund for Scientific Research-FNRS
(Belgium), by IISN-Belgium and by ``Communaut\'e fran\c caise de
Belgique - Actions de Recherche Concert\'ees''.

\def\cprime{$'$}
\providecommand{\href}[2]{#2}\begingroup\raggedright\endgroup

\end{document}